# On the link between atmospheric cloud parameters and cosmic rays

J. CHRISTODOULAKIS[1], C. A. VAROTSOS[1]*, H. MAVROMICHALAKI[2] and M. N. EFSTATHIOU[1]

[1]*Department of Environmental Physics and Meteorology,* National and Kapodistrian *University of Athens, Athens, GR*
[2]*Department of Nuclear and Particle Physics, National and Kapodistrian University of Athens GR*

**Abstract.** We herewith attempt to investigate the cosmic rays behavior regarding the scaling features of their time series. Our analysis is based on cosmic ray observations made at four neutron monitor stations in Athens (Greece), Jung (Switzerland) and Oulu (Finland), for the period 2000 to early 2017. Each of these datasets was analyzed by using the Detrended Fluctuation Analysis (DFA) and Multifractal Detrended Fluctuation Analysis (MF-DFA) in order to investigate intrinsic properties, like self-similarity and the spectrum of singularities. The main result obtained is that the cosmic rays time series at all the neutron monitor stations exhibit positive long-range correlations (of 1/f type) with multifractal behavior. On the other hand, we try to investigate the possible existence of similar scaling features in the time series of other meteorological parameters which are closely associated with the cosmic rays, such as parameters describing physical properties of clouds.

## 1. Introduction

Cosmic rays (CR) illustrate the radiation of particles coming from stellar sources inside or, outside the solar system, consisting mainly of high energy protons (~ 89%), alpha particles (~ 10%) and other heavier cores (~ 1%). In the case, CR originate outside the solar system they are characterized as Galactic Cosmic Rays (GCRs). In this regard, high-energy astrophysical processes, such as the supernova, are believed to produce most of the GCRs traveling in the universe [1].

When CR reach the Earth's atmosphere, then showers of muons, electrons, neutrinos, gammas, positrons, neutrons, protons, p+, K+ (i.e. secondary particles) are produced, penetrating deeper in atmosphere and, depending on their energies, reach the Earth's surface where they are monitored by ground-based detectors. The inventor of the Neutron Monitor was John A. Simpson, since 1948 [2]. In 1965 J. Simpson, along with his students and co-workers, built the first cosmic ray (energy particle) detectors to visit Mars. Simpson was one of the 12 scientists, who organised the programme of the 1957-1958 International Geophysical Year (IGY), to study cosmic rays, solar physics, and magnetospheric physics, when many stations, of the present-day worldwide network, the continuous monitoring started.

It is worth noting that neutron monitor stations, like the ones used in this study, are observing the hadronic component of the secondary CR. These measurements of secondary radiation offer an important tool for the study of primary CR as they are directly correlated with each other [3].

Several studies presented their analytical results about the temporal evolution of CR. For example, [4] studied the temporal evolution of cosmic ray daily values based on the dataset collected by neutron monitors at different cutoff rigidities, by applying a wavelet transform tool (time scale ~60 to ~1000 days). However, this analysis did not lead to a persistent periodicity with the same amplitude for the entire period analyzed.

In this regard, [5] investigated the long-term modulation of the GCRs over the past 1150 years, using the $^{10}$Be data recorded at Greenland and the South Pole and introducing the use of 22-year average intensity of GCR. Greenland data, due to their high temporal resolution, indicated that there were significant 11-year, and also other, fluctuations superimposed upon the high GCR intensities during the Spoerer and Maunder solar activity minima.

These findings have demonstrated the continued presence of an effective and time-dependent heliomagnetic field. [5] also stressed the point that the modulation (i.e., depression) of the cosmic ray intensity during the instrumental era (1933–present) was one of the greatest in the past 1150 years. They, interpreting their results, concluded also that the long-term variations in the GCR intensity are poorly related to sunspot number during periods of low solar activity. Finally, [5] showed that there is a relatively good correlation between variations in the $^{10}$Be data and changes in the open solar magnetic flux predicted by the [6] and [7] models, but there are significant differences among these models and $^{10}$Be data over specific periods.

In addition, [8] have recently proposed a new mechanism for geomagnetic field - GCR influence on the near tropopause ozone, notably: (i) the higher level of Pfotzer maximum (placed above the tropopause) could explain the higher density of the lower stratospheric ozone in regions with stronger geomagnetic field – due to the environmental conditions favouring activation of autocatalytic ozone production; (ii) in regions with a weaker geomagnetic field the ionisation layer (formed by GCR) is placed in the upper troposphere, whose greater humidity stimulates the activation of ozone destructive chemistry. This mechanism needs further exploration for better understanding of the ozone variability which is of great interest in both air-pollution and ozone layer depletion environmental problems and their impacts [9-22].

The present study aims to study the temporal evolution of cosmic rays, along with the intrinsic self-similarity and the spectrum of singularities in their time series, using CR intensity daily values of neutron monitor stations obtained through the Neutron Monitor DataBase (NMDB) and also investigate for plausible connections with cloud parameters.

## 2. Data and Analysis

NMDB (http://www.nmdb.eu/) was founded 10 years ago under the European Union's FP7 program. It is known for its successful and continuous operation, distributing officially data of the neutron monitor stations for several applications. For the purpose of the present study the data sets of four of these stations, located at Athens (Greece), at Jung (Swiss) and at Oulu (Finland), were used (http://www.nmdb.eu/nest/).

Athens NEutron MOnitor Station (A.NE.MO.S) (37.97° N, 23.78° E, altitude: 260m asl, Effective vertical cutoff rigidity: 8.53GV) initiated its activity in November 2000. From the website of the station (http://cosray.phys.uoa.gr) 1-min and 1-hour data are available online, while the resolution of the measurements reaches up to 1-sec. The Athens Neutron MOnitor DAta Processing (ANMODAP) Center has been established and operating at the A.NE.MO.S since 2003. In ANMODAP Center network initially participated 23 NM stations while now these stations have been increased to 34. Data from the network are gathered in real time along with satellite observations from the Advanced Composition Explorer (ACE) and the Geostationary Operational Environmental Satellite (GOES) [23].

Jung IGY NM station (IGYNMs) (46.55° N, 7.98° E, altitude: 3570m asl, Effective vertical cutoff rigidity: 4.5GV) and NM64 NM station (NM64NMs) (46.55° N, 7.98° E, altitude: 3475m asl, Effective vertical cutoff rigidity: 4.5GV) initiated their activities in October 1958 and January 1986, respectively. They are two of the oldest NM stations with continuous operation worldwide. The operation of both stations is supported by the Physikalisches Institut of the University of Bern and by the International Foundation High Altitude Research Stations Jungfraujoch and Gornergrat (HFSJG) in Bern.

Oulu NM station (OULU) (65.05° N, 25.47° E, altitude: 15m asl, Geomagnetic cutoff: 0.8GV) initiated its activity in April 1964. It is operated by Sodankyla Geophysical Observatory of the University of Oulu, Finland.

The daily cosmic rays data obtained at the Athens, Jung and Oulu stations are used in the present analysis, after their corrections for pressure and efficiency over the periods 10/11/2000 – 31/03/2017 for A.NE.MO.S and 01/01/2000 – 31/03/2017 for IGYNMs, NM64NMs and OULU.

To investigate the existence of scaling dynamics in the mentioned time series, we employed the DFA2 technique, which eliminates the noise of non-stationary time series and detects their scaling features [24-33].

The main idea behind DFA technique is the removal of seasonal variations and non-stationarities from a time series by dividing it into different segments of equal length, $\tau$, and studying among segments scaling dynamics characteristics. It provides a relationship between the root mean square fluctuations $F_d(\tau)$ and the segment size $\tau$, characterized for a power-law $F_d(\tau) \propto \tau^\alpha$. The exponent $\alpha$ is the scaling exponent, which is basically a self-affinity parameter indicating the long-range power-law correlation properties of the time-series (fractal properties). The value $\alpha = 0.5$ represents white noise, $\alpha < 0.5$ indicates antipersistent long range correlations, and $\alpha > 0.5$ characterizes persistent long range correlations. A detailed description of the DFA sequential steps is given in [34], as well as in [35].

Additionally, we used the Multifractal Detrended Fluctuation Analysis (MF-DFA2) in order to examine the singularity spectrum of the cosmic rays time series and to estimate the multifractality degree. The generalized MF-DFA procedure was proposed by [36] and its sequential steps are described, in details, in [36].

The annual cycle that characterizes the cosmic rays time series was removed by using the average values of $CR_{mean}$ for each calendar day, which were calculated over the entire study period (2000-2017) and subtracted from the corresponding CR time series of the relevant day. On the other side, the long-term trend of CR time series was removed using the polynomial (of $6^{th}$ degree) regression analysis. The finally obtained CR time series is noted as $CR_{net}$.

However, the existence of long-range correlations in the $CR_{net}$ time series was established by employing the autocorrelation function and the method of the local slopes of the fluctuation functions (i.e. the two criteria proposed by [37]).

Finally, we attempted a comparison between $CR_{net}$ time series and the time series of two meteorological parameters describing physical properties of clouds, such as cloud optical thickness liquid mean and cirrus reflectance mean. Daily data of both parameters obtained from the NASA Giovanni website (http://disc.sci.gsfc.nasa.gov/giovanni) were used covering the period July 2002 to March 2017. These measurements were retrieved from the Moderate Resolution Imaging Spectroradiometer (MODIS) on the Terra research satellite (1° lat × 1° long gridded).

## 3. Discussion and Results

For the purposes of the present study the first step was to investigate the possible existence of self-similarity in the CR time series, for the periods 10/11/2000 – 31/03/2017 at A.NE.MO.S and 01/01/2000 – 31/03/2017 at IGYNMs, NM64NMs and OULU. The results obtained from the analysis performed in the time series of the cosmic rays data are presented for each station just below.

### 3.1 The case of the Athens Neutron Monitor Station

The $CR_{net}$ time series at A.NE.MO.S is shown in Fig.1 (left panel), and the corresponding root-mean-square fluctuation function $F_d(\tau)$ versus time scale $\tau$ (in days), is depicted in Fig.1 (right panel).

The scaling exponent extracted from the DFA2 application on the above described time series was found $\alpha = 1.08 \pm 0.01$, suggesting long-range persistence (of $1/f$ – type). However, the establishment of the power-law long-range correlations in the $CR_{net}$ time series requires the investigation of the rejection of the exponential decay of the autocorrelation function and the constancy of "local slopes" in a certain range towards the low frequencies [37].

In more details, since the single straight line of the DFA2 plot for the $CR_{net}$ time series (at A.NE.MO.S) detected in the entire range of scales ($\alpha = 1.08$) might be biased, we evaluated the local slopes of $\log F_d(\tau)$ vs. $\log\tau$ (separately for two different window sizes of 15 and 20 points, which were shifted successively over all the calculated scales $\tau$), asking for constancy in a sufficient range. To this aim, we performed Monte Carlo simulations applying the DFA2 method on 500 times series characterized by fractional Gaussian noise (with $\alpha = 1.08$, sufficient constancy of local slopes and power-law scaling in the autocorrelation function for a wide range of scales) in order to calculate the local slopes-$\alpha(\tau)$ for each of the 500 time series, at a window of 15 points that was shifted successively over all the calculated scales $\tau$ [38].

It is worthy of note that the Kolmogorov–Smirnov [39] and Anderson–Darling [40] best fit tests showed that, for a fixed scale $\tau$, the dataset of the derived local slopes-$\alpha(\tau)$ obey Gaussian distribution (at 95% confidence level). As estimates of the 95% confidence bands we considered $\alpha(\tau) \pm 2\sigma_{\alpha(\tau)}$.

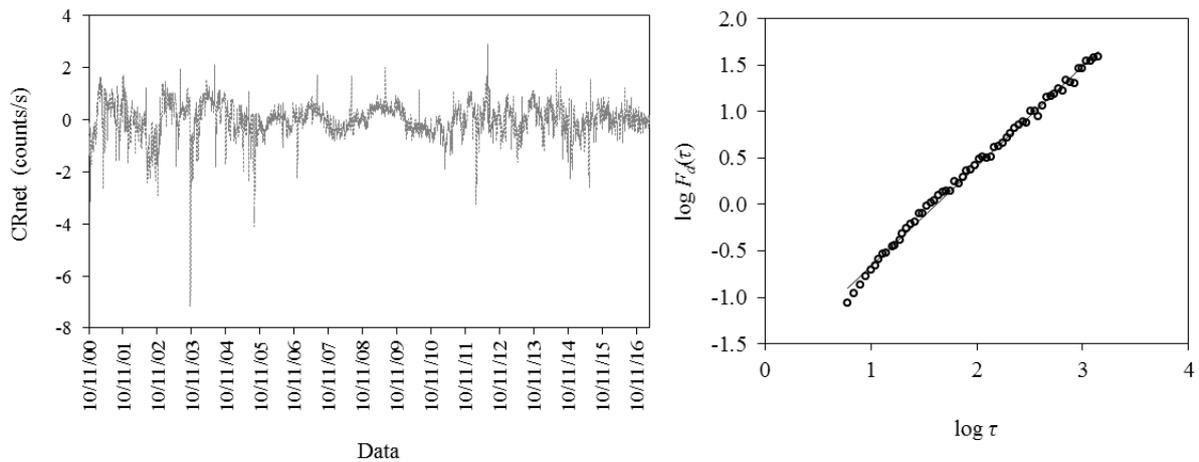

**Fig. 1.** Time series of $CR_{net}$ daily values (after removing long-term trend and annual cycle), during the period 10/11/2000 – 31/03/2017 at A.NE.MO.S (left panel). The corresponding root-mean-square fluctuation function $F_d(\tau)$ of DFA2 versus time scale $\tau$ (in days), in log-log plot and the respective best fit equation ($y = 1.08x - 1.74$, with $R^2 = 0.996$) (solid line, right panel).

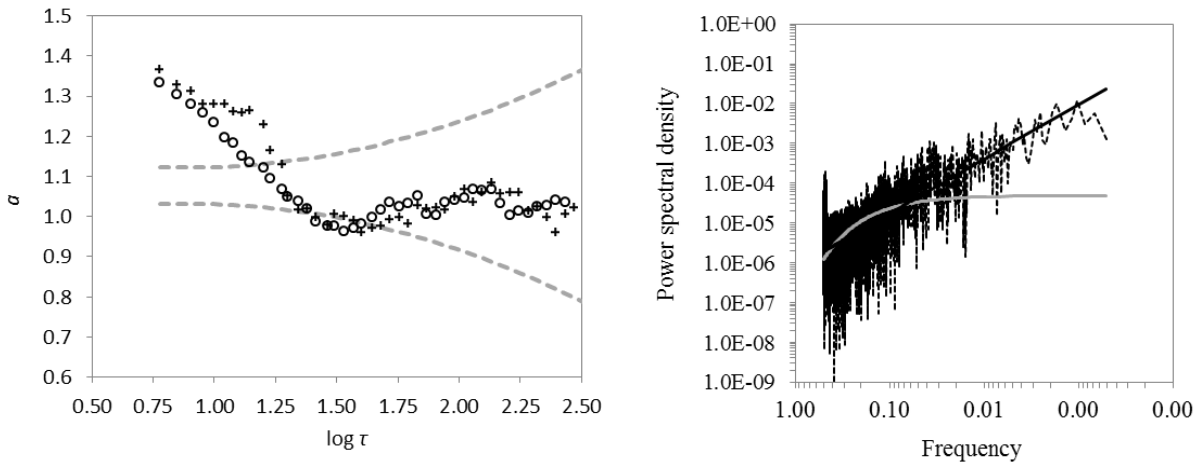

**Fig. 2.** Local slopes of the $\log F_d(\tau)$ vs. $\log\tau$ (10-base logarithms) calculated within a window of 15 points (crosses +) and of 20 points (circles o) for the $CR_{net}$ time series at A.NE.MO.S. The dashed grey line indicates the corresponding $2\sigma$ intervals around the mean value of local slopes ($\alpha = 1.08$) (left panel). Power spectral density for the above mentioned time series, with the corresponding power-law (grey solid line) and the exponential (black solid line) fit ($y = 7.34 \cdot 10^{-7} x^{-1.37}$, with $R^2 = 0.45$ and $y = 4.84 \cdot 10^{-5} e^{-7.39x}$, with $R^2 = 0.32$) (right panel).

In the following, we attempted to determine a range $R$ for the $a$-local slopes of the CR$_{net}$ time series, based on their mean value increased and decreased by the two standard deviation $s_{a(\tau)}$, as derived from the 500 estimated local slopes- $\alpha(\tau)$, i.e. $R = (\alpha - 2\sigma_{\alpha(\tau)}, \alpha + 2\sigma_{\alpha(\tau)})$. Thus, according to Fig. 2 (left panel), all the local slopes (after the scale $\log\tau=1.8$) lie within the borders range $R$, suggesting a sufficient constancy.

Fig. 2 (right panel) presents the profile of the power spectral density for the CR$_{net}$ time series at A.NE.MO.S, showing that the power-law fit gives higher coefficient of determination compared to the exponential fit. Thus, both criteria of [37] are satisfied and therefore the long-range correlations of power-law type for the CR time series at the Athens station are established.

Furthermore, to examine whether the aforementioned value of the DFA2-exponent ($\alpha = 1.08$) is attributed to the temporal evolution of CR$_{net}$ values and not from their marginal distribution, the examined time series was randomly shuffled. If the shuffled CR$_{net}$ values followed the random (white) noise, then the persistence found above would not come from the data but from their temporal evolution. To this aim, we applied the Monte Carlo method, relying on 1000 repeated random samplings of the CR$_{net}$ dataset of A.NE.MO.S to estimate the corresponding extracted exponents. According to the Kolmogorov–Smirnov [39] and Anderson–Darling [40] best fit tests, the dataset of the derived $\alpha''$-exponents obeyed Gaussian distribution at 95% confidence level with mean value $\alpha'' = 0.53$ and standard deviation $\sigma_{\alpha''} = 0.08$. Moreover, the 95% confidence interval of $\alpha''$ was $\left(\alpha'' - 1.96\frac{\sigma_{\alpha''}}{\sqrt{1000}}, \alpha'' + 1.96\frac{\sigma_{\alpha''}}{\sqrt{1000}}\right)$. Thus, according to the t-test, the estimated DFA exponents depicted of CR$_{net}$ time series (at A.NE.MO.S), $\alpha = 1.08$, does not obviously belong to 95 % confidence interval of $\alpha''$ (i.e. white noise) certifying thus long-range persistent behavior.

Our next step was to study the spectrum of singularities for the CR$_{net}$ time series, at A.NE.MO.S, employing the MF-DFA2 technique and calculating the $q^{th}$ order fluctuation function $F_q(\tau)$ for various moments $q$. According to Fig. 3 (upper left panel), the scaling behavior of $F_q(\tau)$ (i.e. slope) for all the selected positive (negative) moments $q$ is almost the same for $\log\tau > 2$ ($\tau > 100$), but not for smaller time scales ($\tau < 100$), where the slope of $F_q(\tau)$ increases for less positive (more negative) moments $q$. This behavior indicated the existence of a great degree of multifractality only for smaller time scales ($\tau \leq 100$), a fact that is expected as the large segments cross several local periods with both small and large fluctuations (i.e. negative and positive $q$, respectively) and will therefore average out their differences in magnitude [41].

The above suggested multifractality is established in Fig. 3 (upper right panel) where the generalized Hurst exponent $h(q)$ for the CR$_{net}$ time series (at A.NE.MO.S) varies versus $q$ values (i.e. $h(q)$ is not independent of $q$), while the $h(q)$ values which are higher than 0.5 indicate long-term persistence for the examined time series.

Moreover, the slope of $h(q)$ for positive moments seems to be similar to that one of negative moments, a fact which is in agreement with the findings of Fig. 3 (upper left panel).

Finally, Fig. 3 (lower left panel) presents the singularity spectrum $f(n)$ as a function of the singularity strength $n$ for the examined CR$_{net}$ time series (see [36]). The maximum value of $f(n)$ corresponds to $q = 0$, while $f(n)$ values on the left (right) of the maximum value correspond to positive (negative) moments $q$. It is apparent that $f(n)$ fluctuates similarly on both sides of its maximum value. This fact reveals once more common features of multifractality for positive and negative $q$-values.

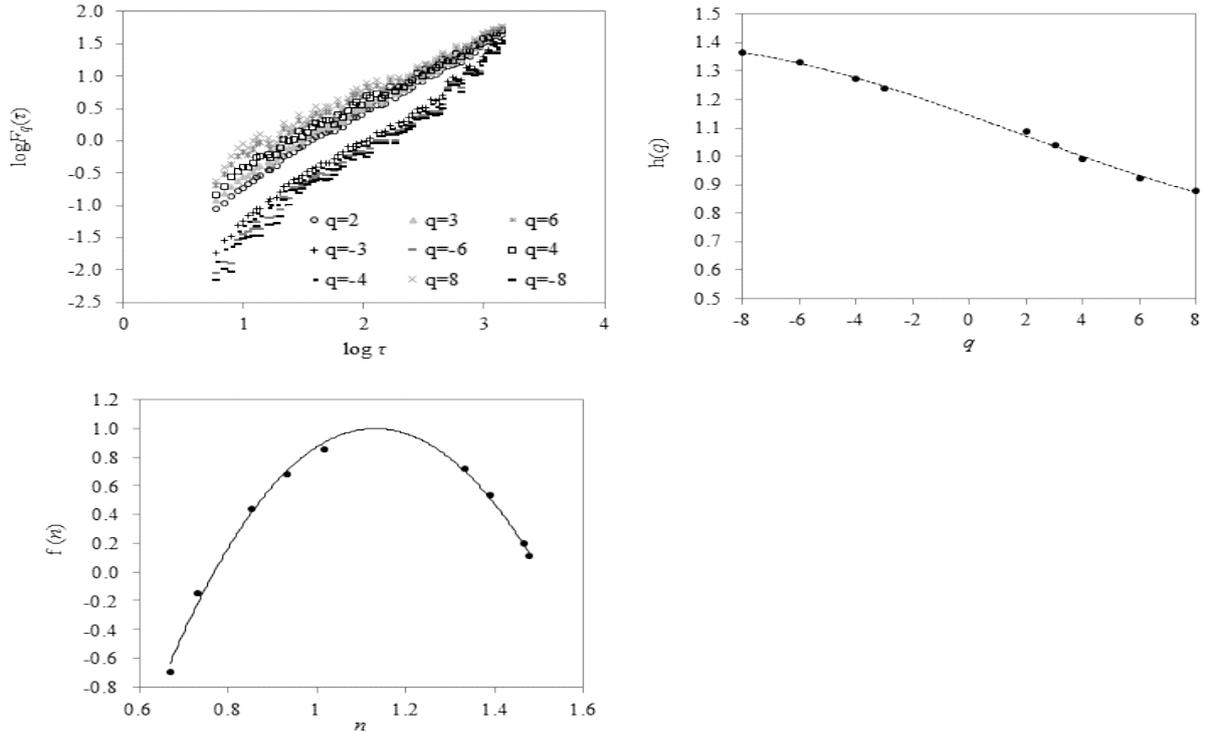

**Fig. 3.** Log-log plots of the MF-DFA2 fluctuation factor $F_q(s)$ versus the time scale $s$ for specific moments $q$ for the $CR_{net}$ time series (at A.NE.MO.S) (upper left panel). Generalized Hurst exponent $h(q)$ versus $q$ for the examined time series. The empirical curve (dots) is fitted by the polynomial of the third order ($y = 8 \cdot 10^{-5} x^3 - 0.0004x^2 - 0.036x + 1.14$, with $R^2 = 0.998$) (dashed line, upper right panel). Singularity spectrum $f(n)$ versus singularity strength $n$ for the examined time series. The empirical curve (dots) is fitted by the polynomial of third order ($y = 0.61x^3 - 9.46x^2 + 19.06x - 9.3$, with $R^2 = 0.993$) (solid line, lower left panel).

### 3.2 The case of the Jung NM64 Neutron Monitor Station

Our next step was to re-apply the above described analysis to the $CR_{net}$ time series at NM64NMs (see Fig. 4, left panel). The corresponding root-mean-square fluctuation functions $F_d(\tau)$ of the DFA2 technique versus time scale $\tau$ (in days), appear in Fig. 4 (right panel).

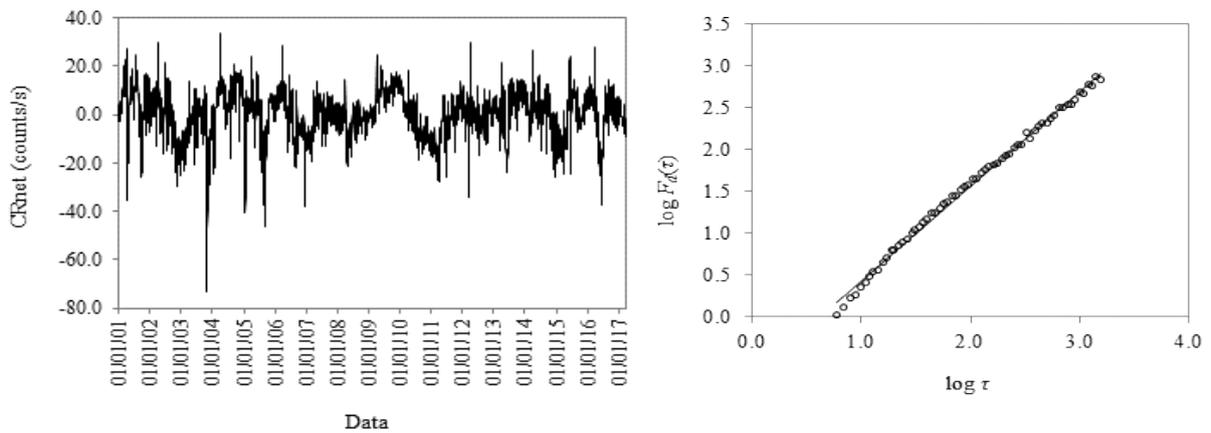

**Fig. 4.** Temporal march of $CR_{net}$ daily values (after removing trend and annual cycle), during the period 01/01/2000 – 31/03/2017 at NM64NMs (left panel). The corresponding root-mean-square fluctuation function $F_d(\tau)$ of the DFA2 versus time scale $\tau$ (in days), in log-log plot and the respective best fit equation ($y = 1.13x - 0.7$, with $R^2 = 0.995$) (solid line, right panel).

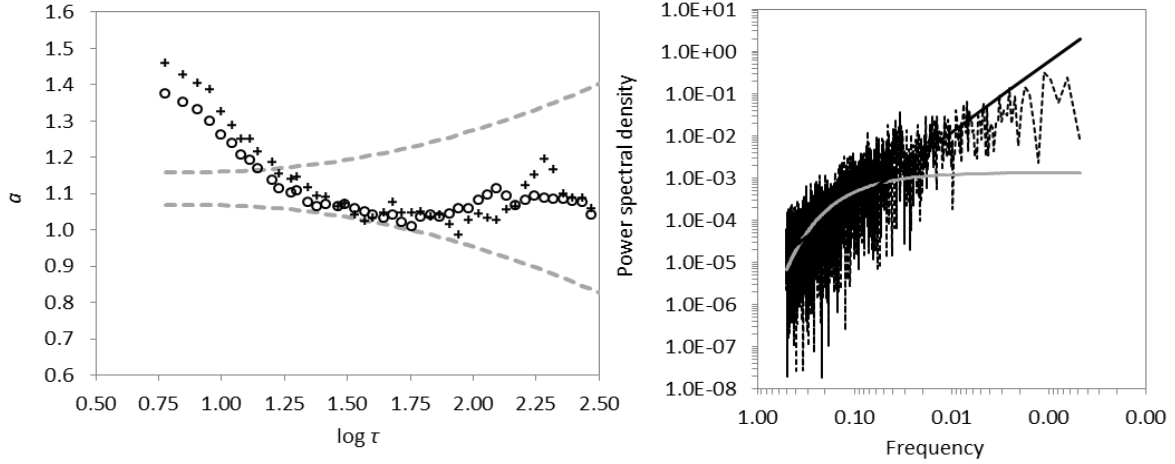

**Fig. 5.** Local slopes of $\log F_d(\tau)$ vs. $\log \tau$ (10-base logarithms) calculated within a window of 15 points (+) and of 20 points (o) for the $CR_{net}$ time series at NM64NMs. The dashed grey line indicates the corresponding $2\sigma$ intervals around the mean value of local slopes ($\alpha = 1.09$) (left panel). Power spectral density for the above mentioned time series, with the corresponding power-law (grey solid line) and the exponential (black solid line) fit ($y = 4.4 \cdot 10^{-5} x^{-1.72}$, with $R^2 = 0.6$ and $y = 1.19 \cdot 10^{-2} e^{-10.8x}$, with $R^2 = 0.51$) (right panel).

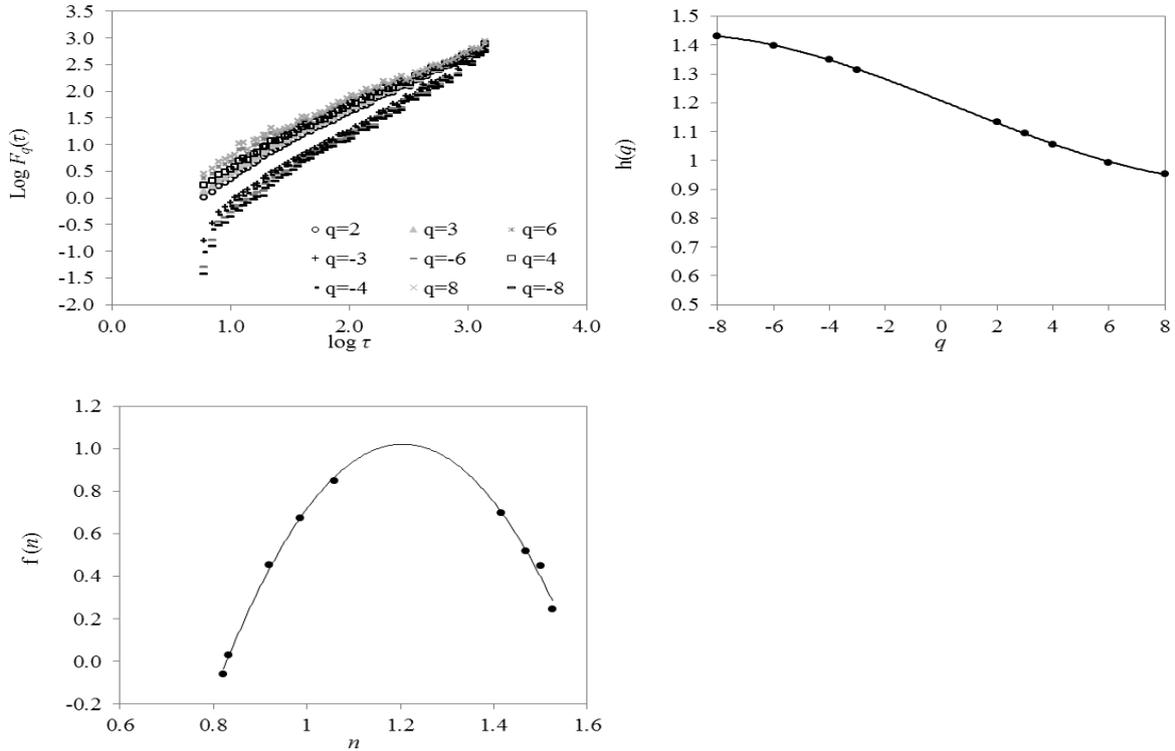

**Fig. 6.** Log-log plots of the MF-DFA2 fluctuation factor $F_q(s)$ versus the time scale $s$ for specific moments $q$ for the $CR_{net}$ time series (at NM64NMs) (upper left panel). Generalized Hurst exponent $h(q)$ versus $q$ for the examined time series. The empirical curve (dots) is fitted by the polynomial of the third order ($y = 10^{-4} x^3 - 0.0002 x^2 - 0.039 x + 1.21$, with $R^2 = 0.999$) (solid line, upper right panel). Singularity spectrum $f(n)$ versus singularity strength $n$ for the examined time series. The empirical curve (dots) is fitted by the polynomial of the third order ($y = -0.033 x^3 - 7.08 x^2 + 17.2 x - 9.37$, with $R^2 = 0.99$) (solid line, lower left).

The scaling exponent revealed from the DFA2 tool for the above described time series is almost equal to that one of the $CR_{net}$ time series at Athens station, suggesting once more persistent memory (of 1/f – type). However, to establish the power-law long-range correlations for the $CR_{net}$ time series (at Jung NM64NMs) we investigated again the satisfaction of the two criteria proposed by [37]. More specifically, we evaluated the local slopes of $\log F_d(\tau)$ vs. $\log \tau$, which belong within the range $R$ over all the calculated scales $\tau$, indicating once more constancy (Fig. 5, left panel).

On the other hand, power-law fit on the profile of the power spectral density for the $CR_{net}$ time series at NM64NMs, seemed to give a better coefficient of determination compared to the exponential fit (Fig. 5, right panel). Thus, both criteria of [37] suggest again long-range correlations of power-law type for the $CR_{net}$ time series at Jung station.

Regarding the spectrum of singularities for the $CR_{net}$ time series, at NM64NMs, we employed again the MF-DFA2 which gave similar results with those over Athens (Fig. 6).

### 3.3 The case of the Jung IGY Neutron Monitor Station

For comparison reasons we examined the case of IGYNMs by re-applying the same analysis to the $CR_{net}$ time series. The corresponding root-mean-square fluctuation functions $F_d(\tau)$ of the DFA2 technique versus time scale $\tau$ (in days) gave scaling exponent ($\alpha = 1.09 \pm 0.01$) indicating again persistent memory (of 1/f – type). Moreover, the two criteria proposed by [37] revealed constancy of local slopes and power-law long-range correlations for the $CR_{net}$ time series (at Jung IGYNMs) (see Table 1).

As far as the spectrum of singularities for the $CR_{net}$ time series, at Jung IGYNMs, we used again the MF-DFA2 which certifying the results which were revealed at A.NE.MO.S and NM64NMs (see Table 1).

### 3.4 The case of the Oulu Neutron Monitor Station

Our last step was the case of OULU station. The application of the DFA2 technique to the $CR_{net}$ time series versus time scale $\tau$ (in days) gave scaling exponent ($\alpha = 1.11 \pm 0.01$) which denotes once more persistency (of 1/f – type). At the same time, the two criteria proposed by [37] revealed again constancy of local slopes and power-law long-range correlations for the $CR_{net}$ time series (at OULU) (see Table 1).

Finally, the MF-DFA2 used to study the spectrum of singularities for the $CR_{net}$ time series, at OULU, gave almost the same results with those ones of A.NE.MO.S, IGYNMs and NM64Nms (see Table 1).

### 3.5 Comparison between cosmic rays and cloud physical parameters

The main result obtained from the above described analysis is that the cosmic rays time series at all the neutron monitor stations exhibit positive long-range correlations (of 1/f type) with multifractal behavior. To this point, we try to investigate the possible existence of similar scaling features in the time series of cloud physical parameters which are, according to some views in literature [42], closely associated with the cosmic rays, such as cirrus reflectance mean (CRM) and cloud optical thickness liquid mean (COTLM). Daily data of both parameters were used covering the period July 2002 to March 2017, for the area of Athens.

Our first step was to apply the non-parametric Spearman's rank test to determine the correlation coefficient $r_s$ between CR values (at A.NE.MO.S) and CRM values as well as between CR values (at A.NE.MO.S) and COTLM values.

**Table 1.** Scaling features of $CR_{net}$ time series at NM64NMs and at OULU

| | DFA2 - exponent | Constancy of "local slopes" | Power spectral density | MF-DFA2 results |
|---|---|---|---|---|
| **$CR_{net}$ (IGYNMs)** | $y = 1.09x - 1.1$<br>$R^2 = 0.99$ | sufficient constancy after $\log\tau = 1.8$ | $y = 5.6 \cdot 10^{-6} x^{-1.67}$ with $R^2 = 0.58$<br>$y = 1.33 \cdot 10^{-3} e^{-10.6x}$ with $R^2 = 0.51$ | • Scaling behavior of $F_q(\tau)$ for all the selected positive (negative) moments $q$ is almost the same for $\tau > 100$, but not for smaller time scales $\tau < 100$, where the slope of $F_q(\tau)$ increases for less positive (more negative) moments $q$.<br>• $h(q)$ is not independent of $q$, while the $h(q)$ values which are higher than 0.5 indicate long-term persistence.<br>• The slope of $h(q)$ for positive moments seems to be similar to that one of negative moments (the polynomial fit of the empirical curve is $y = 4 \cdot 10^{-5}x^3 - 0.0005x^2 - 0.028x + 1.12$, with $R^2 = 0.996$).<br>• The maximum value of $f(n)$ corresponds to $q = 0$ and $f(n)$ fluctuates similarly on both sides of its maximum value (the polynomial fit of the empirical curve is $y = 0.78x^3 - 11.58x^2 + 22.7x - 11$, with $R^2 = 0.994$). |
| **$CR_{net}$ (OULU)** | $y = 1.11x - 1.3$<br>$R^2 = 0.996$ | sufficient constancy after $\log\tau = 1.8$ | $y = 3.08 \cdot 10^{-6} x^{-1.63}$ with $R^2 = 0.58$<br>$y = 6.46 \cdot 10^{-4} e^{-10.4x}$ with $R^2 = 0.52$ | • Scaling behavior of $F_q(\tau)$ for all the selected positive (negative) moments $q$ is almost the same for $\tau > 100$, but not for smaller time scales $\tau < 100$, where the slope of $F_q(\tau)$ increases for less positive (more negative) moments $q$.<br>• $h(q)$ is not independent of $q$, while the $h(q)$ values which are higher than 0.5 indicate long-term persistence.<br>• The slope of $h(q)$ for positive moments seems to be similar to that one of negative moments (the polynomial fit of the empirical curve is $y = 9 \cdot 10^{-5}x^3 - 0.0006x^2 - 0.034x + 1.16$, with $R^2 = 0.997$).<br>• The maximum value of $f(n)$ corresponds to $q = 0$ and $f(n)$ fluctuates similarly on both sides of its maximum value (the polynomial fit of the empirical curve is $y = 2.38x^3 - 15.4x^2 + 25.7x - 11.9$, with $R^2 = 0.99$). |

For the first case, the extracted coefficient was $r_s = -0.016$ suggesting thus that the hypothesis $H_o$: $r_s = 0$ vs $H_1$: $r_s \neq 0$ might be accepted at 95% confidence level (its significance was tested using the t-test). Similarly, for the second case, the derived coefficient was $r_s = -0.03$ indicating a very week anti-correlation at 95% confidence level, while the hypothesis $H_o$ might be accepted at 99% confidence level.

Our next step was to apply the above mentioned scaling analysis on the time series of both cloud physical parameters CRM and COTLM. Table 2 shows the extracted DFA2 exponents ($\alpha = 0.56 \pm 0.01$ for CRM and $\alpha = 0.59 \pm 0.01$ for COTLM) which denote week persistent memory (but not of $1/f$ – type). Moreover, the two criteria proposed by [37] revealed constancy of local slopes for both time series and marginal rejection of the exponential fit only for the COTLM power spectral density (see Table 2).

**Table 2.** Scaling features of CRM and COTLM time series.

| | **DFA2 - exponent** | **Constancy of "local slopes"** | **Power spectral density** | **MF-DFA2 results** |
|---|---|---|---|---|
| **CRM** | $y = 0.76x - 2.01$<br>$R^2 = 0.99$ | sufficient constancy after $\log\tau = 1.5$ | $y = 3.2 \cdot 10^{-7} \cdot x^{-0.24}$ with $R^2 = 0.28$<br>$y = 7.3 \cdot 10^{-7} \cdot e^{-1.68x}$ with $R^2 = 0.29$ | • Scaling behavior of $F_q(\tau)$ for all the selected positive moments $q$ is almost the same for $\tau > 100$, but not for smaller time scales $\tau < 100$, where the slope of $F_q(\tau)$ increases. Similar results were extracted for the selected negative moments $q$.<br>• $h(q)$ values are higher than 0.5 suggesting long-term persistence.<br>• The slope of $h(q)$ for positive moments seems to be a little different from that one of negative moments (the polynomial fit of the empirical curve is $y = 3 \cdot 10^{-4} x^3 + 9 \cdot 10^{-4} x^2 - 0.056x + 0.69$, with $R^2 = 0.997$).<br>• The maximum value of $f(n)$ corresponds to $q = 8$ *revealing different features of multifractality for positive and negative q-values* (the polynomial fit of the empirical curve is $y = 15.8x^3 - 49.6x^2 + 45.66x - 10.9$, with $R^2 = 0.89$). |
| **COTLM** | $y = 0.71x + 0.24$<br>$R^2 = 0.99$ | sufficient constancy after $\log\tau = 1.5$ | $y = 6.1 \cdot 10^{-3} \cdot x^{-0.28}$ with $R^2 = 0.41$<br>$y = 1.5 \cdot 10^{-2} \cdot e^{-1.69x}$ with $R^2 = 0.32$ | • Scaling behavior of $F_q(\tau)$ for all the selected positive moments $q$ is almost the same for $\tau > 100$, but not for smaller time scales $\tau < 100$, where the slope of $F_q(\tau)$ increases. Similar results were extracted for the selected negative moments $q$.<br>• $h(q)$ values are higher than 0.5 suggesting long-term persistence.<br>• The slope of $h(q)$ for positive moments seems to be a little different from that one of negative moments (the polynomial fit of the empirical curve is $y = 3 \cdot 10^{-4} x^3 + 9 \cdot 10^{-4} x^2 - 0.056x + 0.69$, with $R^2 = 0.997$).<br>• The maximum value of $f(n)$ corresponds to $q = 8$ *revealing different features of multifractality for positive and negative q-values* (the polynomial fit of the empirical curve is $y = 6.02x^3 - 24.8x^2 + 24.4x - 5.99$, with $R^2 = 0.97$). |

As far as the spectrum of singularities for the COTLM and CRM time series, we used again the MF-DFA2 technique which gave a little different results from those ones of $CR_{net}$ time series at A.NE.MO.S (see Table 2). In more details, scaling behavior of $F_q(\tau)$ for all the selected positive (negative) moments $q$ is almost the same for $\tau > 100$, but not for smaller time scales $\tau < 100$, where the slope of $F_q(\tau)$ increases for less positive (more negative) moments $q$ (indicating multifractality for smaller time scales). Also, $h(q)$ values are higher than 0.5 suggesting long-term persistence. However, the slope of $h(q)$ for positive moments seems to be a little different from that one of negative moments, as well as the maximum value of $f(n)$ corresponds to $q = 8$ and $f(n)$ doesn't fluctuate similarly on the two sides of its maximum value. This fact reveals different features of multifractality for positive and negative $q$-values. These findings seem to agree with [43] where, rejecting the suggestion of [42], claim that the observed changes in COTLM are not causally related with cosmic rays variations.

## 4. Conclusions

We investigated the intrinsic self-similarity and the spectrum of singularities of cosmic rays time series at four stations, A.NE.MO.S, IGYNMs, NM64Nms and OULU, for the period 2000-2017. The technique employed was DFA2 and MF-DFA2. The main conclusion obtained is that cosmic rays temporal evolution exhibits positive long-range correlations of 1/f type and multifractal behavior, which were detected at all monitoring stations.

In particular, the application of MF-DFA2 technique showed a great degree of multifractality only for smaller time scales ($s \leq 100$), which is expected as the large segments cross several local periods with both small and large fluctuations (i.e. negative and positive $q$, respectively) and will therefore average out their differences in magnitude. However, common features of multifractality were revealed for both positive and negative $q$-values.

In the following, we compared $CR_{net}$ time series at A.NE.MO.S with the time series of two cloud physical parameters, without however detecting any significant correlation. From the other hand, studying their scaling properties, a week persistent memory (but not of 1/f – type) was derived for both COTLM and CRM time series and MF-DFA2 revealed different features of multifractality for positive and negative $q$-values for each cloud parameter.

The results obtained may enhance the fidelity of the sophisticated models not only for the dynamics of the cosmic rays, but also to the study of geophysical and solar parameters that are closely associated with cosmic rays [44-48]. For example the above-mentioned results might give more information about the claimed difference in the solar activity evolution during odd and even solar activity cycles [49].

In addition, the results obtained confirm that NM stations' specifications like their geographical location and effective vertical cutoff rigidity do not affect intrinsic properties of the recorded data like self-similarity and spectrum of singularities. Based on this conclusion, we can argue that for such a kind of statistical study of the CR time series the use of a data set originating from a particular station can lead to conclusions expressing the overall intrinsic characteristics of CR.

## Acknowledgements

We acknowledge the NMDB database (www.nmdb.eu), founded under the European Union's FP7 programme (contract no. 213007) for providing data. Athens neutron monitor data were kindly provided by the Physics Department of the National and Kapodistrian University of Athens. Jung IGY neutron monitor data were kindly provided by the Sphinx Observatory Jungfraujoch. This research did not receive any specific grant from funding agencies in the public, commercial, or not-for-profit sectors.

## References

[1] Jackman, C.H., Marsh, D.R., Kinisson, D.E., Mertens, C.J. & Fleming, E.L. (2016). Atmospheric changes caused by galactic cosmic rays over the period 1960–2010. Atmos. Chem. Phys., 16, 5853–5866.
[2] Simpson, J.A. (2000). The cosmic ray nucleonic component: The invention and scientific uses of the neutron monitor. Space Sci. Rev., 93, 11–32.
[3] Paschalis, P., Mavromichalaki, H., Dorman, L.I., Plainaki, C. & Tsirigkas, D. (2014). Geant4 software application for the simulation of cosmic ray showers in the Earth's atmosphere. New Astron., 33, 26-37.

[4] Kudela, K., Storini, M., Antalova, A. & Rybák, J. (2001). On the wavelet approach to cosmic ray variability. Proceedings of the 27th International Cosmic Ray Conference. 07-15 August, 2001. Hamburg, Germany. Under the auspices of the International Union of Pure and Applied Physics (IUPAP), p.3773.
[5] McCracken, K.G., McDonald, F.B., Beer, J., Raisbeck, G. & Yiou, F. (2004). A phenomenological study of the long-term cosmic ray modulation, 850–1958 AD. J. Geophys. Res. – Space, 109, A12103.
[6] Solanki, S.K., Schussler, M. & Fligge, M. (2002). Secular variation of the Sun's magnetic flux. Astron. Astrophys., 383, 706–712.
[7] Schrijver, C.J., DeRosa, M.L. & Title, A.M. (2002). What is missing from our understanding of long-term solar and heliospheric activity?, Astrophys. J., 577, 1006–1012.
[8] Kilifarska, N.A., Bakhmutov, V.G. & Melnyk, G.V. (2017). Galactic cosmic rays and tropical ozone asymmetries. Comptes rendus de l'Académie bulgare des Sciences, 70(7), 1003–1010.
[9] Tzanis, C., Varotsos, C., & Viras, L. (2008). Impacts of the solar eclipse of 29 March 2006 on the surface ozone concentration, the solar ultraviolet radiation and the meteorological parameters at Athens, Greece. Atmos. Chem. Phys., 8(2), 425-430.
[10] Reid, S.J., Rex, M., Von Der Gathen, P., Fløisand, I., Stordal, F., Carver, G.D., et al. (1998). A Study of Ozone Laminae Using Diabatic Trajectories, Contour Advection and Photochemical Trajectory Model Simulations. J. Atmos. Chem., 30(1), 187-207.
[11] Varotsos, C., Efstathiou, M. & Tzanis, C. (2009). Scaling behaviour of the global tropopause. Atmos. Chem. Phys., 9(2), 677-683.
[12] Varotsos, C. & Cartalis, C. (1991). Re-evaluation of surface ozone over Athens, Greece, for the period 1901–1940. Atmos. Res., 26(4), 303-310.
[13] Tidblad, J., Kucera, V., Ferm, M., Kreislova, K., Brüggerhoff, S., Doytchinov, S., et al. (2012). Effects of air pollution on materials and cultural heritage: ICP materials celebrates 25 years of research. International Journal of Corrosion, Volume 2012 (2012), Article ID 496321, 16 pages.
[14] Varotsos, C., Tzanis, C. & Cracknell, A. (2009). The enhanced deterioration of the cultural heritage monuments due to air pollution. Environ. Sci. Pollut. R., 16(5), 590-592.

[15] Tzanis, C., Varotsos, C., Christodoulakis, J., Tidblad, J., Ferm, M., Ionescu, A., Lefevre, R.-A., Theodorakopoulou, K. & Kreislova, K. (2011). On the corrosion and soiling effects on materials by air pollution in Athens, Greece. Atmos. Chem. Phys., 11 (23), 12039-12048.
[16] Tzanis,C., Varotsos, C., Ferm, M., Christodoulakis, J., Assimakopoulos, M.N. & Efthymiou, C. (2009). Nitric acid and particulate matter measurements at Athens, Greece, in connection with corrosion studies. Atmos. Chem. Phys., 9 (21), 8309-8316.
[17] Feretis, E., Theodorakopoulos, P., Varotsos, C., Efstathiou, M., Tzanis, C., Xirou, T., Alexandridou, N. & Aggelou M. (2002). On the plausible association between environmental conditions and human eye damage. Environ. Sci. Pollut. R., 9 (3), 163-165.
[18] Krapivin, V.F., Varotsos, C.A. & Soldatov, V.Y. (2015). New Ecoinformatics Tools in Environmental Science. Springer International Publishing, Series Title: Environmental Earth Sciences, pp. 903, doi: 10.1007/978-3-319-13978-4.
[19] Cracknell, A.P., Krapivin, V.F. & Varotsos, C.A. (2009). Global Climatology and Ecodynamics: Anthropogenic Changes to Planet Earth. Springer-Verlag Berlin Heidelberg, Series Title: Environmental Sciences, pp. 518, doi: 10.1007/978-3-540-78209-4.
[20] Cracknell, A.P. & Varotsos, C.A. (2007). Editorial and cover: Fifty years after the first artificial satellite: from Sputnik 1 to ENVISAT. Int. J. Remote Sens., 28 (10), 2071-2072.
[21] Varotsos, C. (2003). What is the lesson from the unprecedented event over Antarctica in 2002? Environ. Sci. Pollut. Res., 10(2), 80–81.
[22] Varotsos, C. (2002). The southern hemisphere ozone hole split in 2002. Environ. Sci. Pollut. Res., 9(6), 375–376.

[23] Mavromichalaki, H., Gerontidou, M., Mariatos, G., Papailiou, M., Papaioannou, A., Plainaki, C., Sarlanis, C. & Souvatzoglou, G. (2009). Athens Neutron Monitor Data Processing Center – ANMODAP Center. Adv. Space Res., 44, 1237-1246.
[24] Efstathiou, M.N. & Varotsos, C.A. (2010). On the altitude dependence of the temperature scaling behaviour at the global troposphere. Int. J. Remote Sens., 31, 343-349.
[25] Peng, C.-K., Buldyrev, S.V., Havlin, S., Simons, M., Stanley, H.E. & Goldberger, A.L. (1994). Mosaic organization of DNA nucleotides. Phys. Rev. E., 49, 1685–1689.
[26] Varotsos, C. (2005). Power-law correlations in column ozone over Antarctica. Int. J. Remote Sens., 26, 3333-3342.
[27] Varotsos, C., Ondov, J. & Efstathiou, M. (2005). Scaling properties of air pollution in Athens, Greece and Baltimore, Maryland. Atmos. Environ., 39, 4041 - 4047.
[28] Varotsos, C.A., Ondov, J.M., Cracknell, A.P., Efstathiou, M.N. & Assimakopoulos, M.-N. (2006). Long-range persistence in global Aerosol Index dynamics. Int. J. Remote Sens., 27, 3593– 3603.
[29] Varotsos, C., Assimakopoulos, M.-N. & Efstathiou, M. (2007). Technical Note: Long-term memory effect in the atmospheric $CO_2$ concentration at Mauna Loa. Atmos. Chem. Phys., 7, 629–634.
[30] Weber, R.O. & Talkner, P. (2001). Spectra and correlations of climate data from days to decades. J. Geophys. Res., 106, 20131–20144.
[31] Varotsos, C., Efstathiou, M. & Cracknell, A.P. (2013). Plausible reasons for the inconsistencies between the modeled and observed temperatures in the tropical troposphere. Geophys. Res. Lett. 40, 4906-4910.
[32] Varotsos, C.A., Efstathiou, M.N. & Cracknell, A.P. (2013). On the scaling effect in global surface air temperature anomalies. Atmos. Chem. Phys. 13, 5243-5253.
[33] Cracknell, A. P., & Varotsos, C. A. (2011). New aspects of global climate-dynamics research and remote sensing. Int. J. Remote Sens., 32(3), 579-600.
[34] Efstathiou, M.N., Tzanis, C., Cracknell, A. & Varotsos, C.A. (2011). New features of land and sea surface temperature anomalies. Int. J. Remote Sens. 32, 3231-3238.
[35] Varotsos, C., Efstathiou, M. & Tzanis, C. (2009). Scaling behaviour of the global tropopause. Atmos. Chem. Phys., 9, 677-683.
[36] Kantelhardt, J.W., Zschiegner, A., Koscielny-Bunde, E., Havlin, S., Bunde, A. & Stanley, H.E. (2002). Multifractal detrended fluctuation analysis of nonstationary time series. Physica A. 316, 87-114.
[37] Maraun, D., Rust, H.W. & Timmer, J. (2004). Tempting long-memory – on the interpretation of DFA results. Nonlinear Proc. Geoph. 11, 495–503.
[38] Timmer, J. & König, M. (1995). On Generating Power Law Noise. Astron. Astrophys., 300, 707–710.
[39] Stephens, M.A. (1974). EDS statistics for goodness of fit and some comparisons. J. Am. Stat. Assoc. 69, 730-737.
[40] Anderson, T.W. & Darling,D.A., (1954). A test of goodness of fit. J. Am. Stat. Assoc., 49, 765-769.
[41] Ihlen, E.A.F. (2012). Introduction to multifractal detrended fluctuation analysis in Matlab. Front. Physiol., 3, 141.
[42] Svensmark, H., Bondo, T. & Svensmark, J. (2009). Cosmic ray decreases affect atmospheric aerosols and clouds. Geophys. Res. Lett. 36, L15101.
[43] Laken,B., Wolfendale, A. & Kniveton, D. (2009). Cosmic ray decreases and changes in the liquid water cloud fraction over the oceans. Geophys. Res. Lett., 36, L23803.
[44] Varotsos, C. (1994). Solar ultraviolet radiation and total ozone, as derived from satellite and ground-based instrumentation. Geophys. Res. Lett., 21(17), 1787-1790.
[45] Schulz, A., Rex, M., Harris, N.R.P., Braathen, G.O., Reimer, E., Alfier, R., Kilbane-Dawe, I., Eckermann, S., Allaart, M., Alpers, M. and Bojkov, B. (2001). Arctic ozone loss in threshold conditions: Match observations in 1997/1998 and 1998/1999. J. Geophys. Res. Atmospheres,

106(D7), 7495-7503.

[46] Efstathiou, M. N., & Varotsos, C. A. (2013). On the 11 year solar cycle signature in global total ozone dynamics. Meteorol. Appl., 20(1), 72-79.

[47] Varotsos, C. A. (1998). Total ozone and solar ultraviolet radiation, as derived from satellite and ground-based instrumentation at Dundee, Scotland. Int. J. Remote Sens., 19(17), 3301-3305.

[48] Kondratyev, K. Y., Varotsos, C. A., & Cracknell, A. P. (1994). Total ozone amount trend at St Petersburg as deduced from Nimbus-7 TOMS observations. Int. J. Remote Sens., 15(13), 2669-2677.

[49] Varotsos, C. A., & Cracknell, A. P. (2004). New features observed in the 11-year solar cycle. Int. J. Remote Sens., 25(11), 2141-2157.